\documentclass[aps,preprint,showpacs,groupedaddress]{revtex4}  
\usepackage{graphicx}  
\usepackage{dcolumn}   
\usepackage{bm}        
\usepackage{amssymb}   

\begin{document}

\title{Steady-state distribution function for a gas of independent electrons far from equilibrium}
\author{Thomas Christen \\ ABB Switzerland Ltd, Corporate Research \\ Baden-D\"attwil, Switzerland}
\date{\today}

\begin{abstract}
The quasi-stationary nonequilibrium distribution function of an independent electron gas interacting
with a medium, which is at local thermal equilibrium, can be obtained by entropy production rate minimization, subject to constraints
of fixed moments.
The result is not restricted to the region near equilibrium (linear response)
and provides a closure of the associated generalized hydrodynamic equations of the electron gas for an arbitrary number of moments. Besides
an access to far from equilibrium states, the approach provides a useful description of semi-classical transport in mesoscopic conductors,
particularly because macroscopic contacts can be naturally taken into account. 
\end{abstract}

\maketitle
{\em Introduction -}
Electron transport in matter is often described by the Boltzmann transport equation (BTE)
for the distribution function $f$, from which the (semi-classical) transport properties
can be calculated \cite{Ziman1967}. Kohler \cite{Kohler1948} proved that the stationary solution of the
{linearized} BTE satisfies a variational principle for the entropy production rate, which
has been widely used to determine
linear transport coefficients \cite{Ziman1967,Kohler1949,Sondheimer1950,Martyushev2006}.
Schlup  \cite{Schlup1975} and Jones \cite{Jones1983} gave arguments for the validity of Kohler's principle
beyond linear response. Note that a {\em linear} BTE does in general not imply a restriction to
the near-equilibrium, i.e., linear response, region. First, provided the linearity is not due to
linearization of a nonlinear BTE, the linear BTE may be valid for large deviations from equilibrium.
Secondly, because forces appear as {\em coefficients} of a term linear in $f$, the
resulting currents are generally nonlinear functions of the forces.   
Below, a method is introduced that provides $f$ as a function of its moments
by minimization of the {total} entropy production rate. The result is not restricted to the near-equilibrium case
(i.e., to linear response), and
serves as a closure of generalized hydrodynamic equations for the moments. 
Recently, it has been shown that an analogous method applied to a photon gas in matter at local thermal equilibrium,
where the BTE is exactly linear, describes nonequilibrium radiation satisfactorily well also far
from equilibrium \cite{ChristenKassubek2009}.\\ \indent
After introducing the method, a few examples will be discussed, including metallic and nonmetallic electric conduction, and
low-frequency transport in mesoscopic conductors.\\ \indent

{\em Basic formulation of the problem -} 
Consider electrons in $d$($=1,2,3$)-dimensional space, with space and velocity vectors ${\bf x}$ and ${\bf v}$, respectively.
They may interact with a medium at local thermal equilibrium,
such that the electron distribution function $f({\bf x}, {\bf v})$  obeys the linear BTE \cite{Ziman1967},
\begin{equation}
\partial _{t} f + {\bf v} \cdot {\bm \nabla } _{x}f + {\bf a}\cdot {\bm \nabla} _{v } f
=\mathcal{L}(f_{0}-f) \;\;.
\label{BTE}
\end{equation}
The force $ m \,{\bf a} = e\,{\bm \nabla} _{x} U $ acting on the particles with effective mass $m$ and electron charge $-e$, is
related to the gradient of the electric potential $U({\bf x})$, which may be determined later (on the hydrodynamic level)
self-consistently from the Poisson equation.
The positive definite and self-adjoint linear (integral-) operator $\mathcal{L}$ describes the interaction of the
electrons with the medium. The terms $- \mathcal{L} f$ and $\mathcal{L} f_{0}$ can be interpreted, respectively, as absorption of electrons and
emission of electrons equilibrated with respect to the medium (e.g., traps, contacts in mesoscopic systems, etc.), which has
local temperature $T $ and (electro-)chemical potential $\mu $.
The index $0$ indicates the Fermi equilibrium distribution $ f_{0} = \{\exp(\frac{H-\mu}{kT})+1\}^{-1}$,
where $H(v)$ is the particle energy, $v= \mid {\bf v} \mid $, and $k$ is the Boltzmann constant.
Equation (\ref{BTE}) applies to a large class of systems in solid state physics \cite{Ziman1967}, presumes micro-reversibility,
and is usually solved within linear response, i.e., in first order of $\delta f = f-f_{0}$. Typical standard methods are
the BGK approximation \cite{BathnagarGrossKrook1954}, or Kohler's variational approach with the help of trial functions.
However, if the linearity is not due to a linearization of a nonlinear scattering term but
remains valid for larger deviations of $f$ from $f_{0}$, a {\em nonlinear (far from equilibrium) solution} is apposite.
Note that particle number conservation is generally not assumed, which allows for
carrier absorption and emission by the medium (see examples below).\\
To derive a quasi-steady state solution for general $f$, an arbitrary number $M+1$ of moments are defined:
\begin{eqnarray}
n ({\bf x})   & = & \left(\frac{m}{h}\right)^{d} \int d^{d}v \, f({\bf x}, {\bf v}) \label{ndef} \\ 
{\bf j} ({\bf x})   & = & \left(\frac{m}{h}\right)^{d} \int d^{d}v \, {\bf v} f({\bf x}, {\bf v}) \label{jdef} \\ 
\Pi _{kl}({\bf x})   & = & \left(\frac{m}{h}\right)^{d} \int d^{d}v \, v_{k} v_{l} f({\bf x}, {\bf v}) \label{Pidef} 
\end{eqnarray}
etc., where $h$ is Planck's constant (spin degeneracy can be included later).
Equations for $M$ moments
follow from multiplication of Eq. (\ref{BTE}) by $1$, $v_{k}$, $v_{k}v_{l}$, ..., and velocity integration:
\begin{eqnarray}
\partial _{t} n + {\bm \nabla} \cdot {\bf j }& = & P^{(n)}\label{nh} \\ 
\partial _{t} {\bf j} + {\bm \nabla} \cdot \Pi - n {\bf a} & = & {\bf P}^{(j)} \;\; \label{jh} 
\end{eqnarray}
etc. ($k,l=1,..., d$). The highest order moment and the terms on the right hand side,
$P^{(n)}=(m/h)^d\int d^d v \mathcal{L}(f_{0}-f)$, ${\bf P}^{(j)}=(m/h)^d\int d^d v {\bf v}\mathcal{L}(f_{0}-f)$, etc., are
functionals of the unknown distribution $f$. In order to close the $M$ moment equations, we will calculate $f$ from
entropy production rate minimization, subject to constraints of $M$ fixed moments,
given by Eqs. (\ref{ndef}), (\ref{jdef}), etc.. The distribution and derived quantities will thus depend on them.\\ \indent
%

{\em Determination of $f$ -}
It is important to consider the {\em total} entropy production rate of the whole isolated system, which consists of two contributions
associated with the independent electron gas and with the medium acting as an equilibrium bath.
The first part, $\dot s_{F}$, can be derived from the entropy density
$s_{F}({\bf x}) =-k \left(\frac{m}{h}\right)^{d} \int d^{d}v \{ f \ln f +(1-f)\ln(1-f) \} $
by differentiation of $s_{F}$ with respect to time, replacement of
$\partial _{t}f $ with the help of Eq. (\ref{BTE}), partial $v$-integration, and finally
writing the result as an entropy balance equation,
$\partial _{t} s_{F} + {\bm  \nabla} \cdot {\bf q}_{s_{F}} =\dot s_{F}$, with entropy current density  ${\bf q}_{s_{F}}$.
This gives $\dot s_{F} = k \left(\frac{m}{h}\right)^{d} \int d^{d}v \;  \ln \left( \frac{f}{1-f} \right)\mathcal{L}(f-f_{0}) $.
The second part, $\dot s_{B}= W/T$, associated with the medium, is obtained from
the heat power density $W=  \left(\frac{m}{h}\right)^{d} \int d^{d}v \;  (H-\mu)\; \mathcal{L}(f-f_{0})$.
The relation $H-\mu = kT \ln (1/f_{0}-1)$ implies
$\dot s_{B} = k \left(\frac{m}{h}\right)^{d} \int d^{d}v \ln \left( \frac{1-f_{0}}{f_{0}} \right)\mathcal{L}(f-f_{0})$.
The total entropy production rate, $\dot s_{B} +\dot s_{F} $, is
\begin{equation}
\dot s =  k \left(\frac{m}{h}\right)^{d} \int d^{d}v \; \ln \left( \frac{f(1-f_{0})}{f_{0}(1-f)}
\right)\mathcal{L}(f-f_{0})  \; .
\label{Totalentropyrate}
\end{equation}
The distribution function $f$ is then determined by the variational principle
$\delta \dot s /\delta f =0$ subject to the constraints Eqs. (\ref{ndef}), (\ref{jdef}), ...
of fixed moments. This is the main result, and provides a closure for the generalized hydrodynamics 
of the independent electron gas far from local equilibrium and for an arbitrary number of moments.\\ \indent

{\em Two-moment closure -} For illustration, consider $M=2$ moments and
assume $ \mathcal{L}(f_{0}-f) = r(v) (f_{0}-f) $ with
$v$-dependent relaxation rate $r$. An example beyond this relaxation-time
approximation will an elastic scatterer in the last example.
Optimization of Eq. (\ref{Totalentropyrate}) with constraints (\ref{ndef}) and (\ref{jdef}) leads to
\begin{equation}
r(v) \left( \ln \frac{f(1-f_{0})}{f_{0}(1-f)}+ \frac{f-f_{0}}{f(1-f)} \right)=
\lambda ^{(n)}+{\bm \lambda} ^{(j)}\cdot {\bf v} \;\;
\label{mainresult}
\end{equation}
with Lagrange parameters $\lambda ^{(n)}$ and ${\bm \lambda} ^{(j)}$. It is readily checked that
the optimization problem is convex, which ensures that the solution is
a constraint minimum.
Solving Eq. (\ref{mainresult}) for $f$ and elimination of the Lagrange
parameters provides $f({\bf v}; n,{\bf j})$ and then $P^{(n)}, {\bf P}^{(j)}$, and $ \Pi_{kl}$.\\ \indent

{\em Weak Nonequilibrium -}
Expansion with respect to $\delta f = f-f_{0}$ gives
$\delta f = f_{0}(1-f_{0})(\lambda ^{(n)}+{\bm \lambda} ^{(j)} \cdot {\bf v})/2r$. Elimination of
$\lambda ^{(n)}$ and ${\bm \lambda} ^{(j)}$ with the help of Eqs. (\ref{ndef}) and (\ref{jdef})
leads to $P^{(n)}= r^{(n)}(n_{0}-n)$ and  ${\bf P}^{(j)} = - r^{(j)}{\bf j}$ with
\begin{eqnarray}
r^{(n)}  & = & \frac{ \int d^{d}v \,f_{0}(1-f_{0})}{ \int d^{d}v \, r^{-1}(v)f_{0}(1-f_{0})} \label{r0} \\
r^{(j)}  & = & \frac{ \int d^{d}v \, v^{2}f_{0}(1-f_{0})}{ \int d^{d}v \, v^{2}r^{-1}(v)f_{0}(1-f_{0})} \label{r} \;\;,
\end{eqnarray}
 For $d=3$, Eq. (\ref{r})
gives the usual linear response relaxation-time mobility, $e/mr^{(j)}$ \cite{Ziman1967}.
The stress tensor, $\Pi _{kl} $, stays diagonal and isotropic,
and deviates from equilibrium $\Pi _{0,kk}$ by  
\begin{equation}
\Delta \Pi _{kk} = \frac{(n-n_{0})}{d}
\frac{\int d^{d}v\; v^{2}r^{-1}f_{0}(1-f_{0})}{\int d^{d}v\; r^{-1}f_{0}(1-f_{0})}\;\;.
\label{Pi1d}
\end{equation}
Note that $ f_{0}(1-f_{0}) = -\frac{kT }{mv}\partial _{v}f_{0} $, and Eqs. (\ref{r0})-(\ref{Pi1d}) simplify in the low
temperature limit for $\mu = mv_{F}^{2}/2 >0$, because $\partial _{v}f_{0} $ is a Dirac $\delta$-function at the equilibrium
Fermi velocity $v_{F}= h(n_{0}/\Omega _{d})^{1/d}/m$ (with $\Omega _{1}=2$, $\Omega _{2}=\pi$, $\Omega _{3}=4\pi /3$).
For $T \to 0$, $ \Pi_{0, kl}=\delta _{kl}n_{0}v_{F}^{2}/(d+2)$.
Finite temperature corrections may be considered in terms of
$\gamma =2kT / mv^{2} _{F}$, in the same way as done by other standard methods for weak nonequilibrium.\\ \indent

\begin{figure}[h]
\centering
    \rotatebox{0}{
        \resizebox{0.3\textwidth}{!}{
            \includegraphics{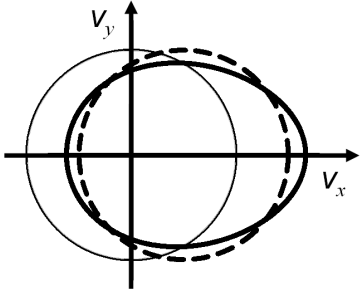}}}
\caption{Far from equilibrium Fermi circle in $2d$ for $n=n_0$ and $j=n_{0}v_{F}/2$ in $x$-direction for constant $r$ (dashed) and deformed
circle for $r\propto 1/v$ (solid). The thin line refers to equilibrium.}
\label{Fermiball}
\end{figure}

{\em Fermi Sphere, Zero Temperature -} 
For $\mu >0$ and $T \to 0$, Eq. (\ref{mainresult}) can be solved analytically for {\em strong deviation from equilibrium}.
After multiplying Eq. (\ref{mainresult}) by $f(1-f)$, using
$(1-f_{0})/f_{0} = \exp (m(v^2-v_{F}^2)/2kT)$ and the fact that $f\in [0,1]$, one can solve the resulting equation
with an expansion $f=f^{(0)}+\gamma f^{(1)}+O(\gamma ^2)$. One concludes for $\gamma  \to 0$ that 
$f^{(0)}=1$ and $f^{(0)}=0$ if $\eta ({\bf v}):=m(v^2-v_{F}^2)/2kT -\lambda ^{(n)}/r -  {\bm \lambda }^{(j)} \cdot {\bf v}
/r $  is negative and positive, respectively (note $\lambda ^{(n,j)}\propto 1/\gamma $). Hence, $\eta ({\bf v})=0$ defines
the boundary of the nonequilibrium Fermi surface for $T\to 0$.
For $v$-independent $r$, the Fermi sphere is shifted according to
${\bf j}/n$ and resized according to $n/n_{0}$ as one expects, and the stress tensor becomes
\begin{equation}
\Pi _{k l} = \frac{h^2}{5m^2} \frac{n^{1+2/d}}{(d+2)\Omega_{d}^{2/d}}\, \delta _{kl} +\, \frac{j_{k}j_{l}}{n} \;\;,
\label{PiNonlin}
\end{equation}
in accordance with the usual expression of the hydrodynamic momentum flux tensor. A factor $2$ in $\Omega _{d}$
must be included for spin degeneracy; for instance, Eq. (\ref{PiNonlin}) leads
then to the $d=3$ electron gas pressure introduced by Bloch \cite{Bloch1933}.\\ 
For $v$-dependent $r$ these results change in general far from equilibrium.
As an example, consider $r(v)=r_{0}v_{F}/v $ with constant $r_{0}$.
For $1d$, one finds $P^{(n)}= n_{0}r_{0}\ln\sqrt{(n/n_{0})^2-(j/nv_{F})^2}$
and $r^{(j)}=r_{0}n_{0}/n$. The divergence of $P^{(n)}$ at $j=v_F n^{2}/n_0$ is an artifact due to
the specific $r(v)$ and occurs when the shifted Fermi `surface' approaches $v=0$.
From $P^{(n)}(n,j)=0$ and Eq. (\ref{jh}) one can derive
the relation between $j$ and the electric field. A simple experimental test of
the theory would be based on a measurement of the nonlinear current-voltage
characteristics of a nano-wire on an appropriate support material, if the associated wire-support interaction in
terms of $r(v)$ is known.\\
A non-spherical deformation of the Fermi sphere for varying current can occur in
$d>1$. For illustration, 
we have calculated the deformed Fermi circle in $d=2$ for $r(v)=r_{0}v_{F}/v $.
By comparing the result (solid thick curve in Fig. \ref{Fermiball}) with a Fermi circle for constant $r$ (dashed curve),
one observes that the former is closer to the equilibrium distribution for low $v$-values
and further away for larger $v$-values. This reflects the stronger
equilibration, induced by the entropy principle \cite{ChristenKassubek2009},
at low $v$ for this $r(v)$. It is clear from Eq. (\ref{Pidef}), that
not only the shift but also the deformation of the Fermi sphere gives a contribution to an anisotropy $\Pi _{kl}$.\\ \indent
%
%

\begin{figure}[h]
\centering
    \rotatebox{0}{
        \resizebox{0.4\textwidth}{!}{
            \includegraphics{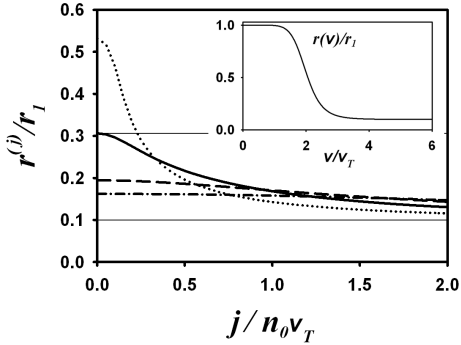}}}
\caption{Mean rate $ r^{(j)} /r_{1}$ as a function of $j$ for $r(v)$ shown in the inset; dotted: $n/n_{0}=0.6$; solid: $n/n_{0}=1$;
dashed: $n/n_{0}=1.4$; dashed-dotted: $n/n_{0}=1.8$. Near equilibrium (upper thin line): $r^{(j)}= 0.306r_{1}$; lower thin line:
$r_2/r_1=0.1$.}
\label{Fig2}
\end{figure}
\begin{figure}[h]
\centering
    \rotatebox{0}{
        \resizebox{0.4\textwidth}{!}{
            \includegraphics{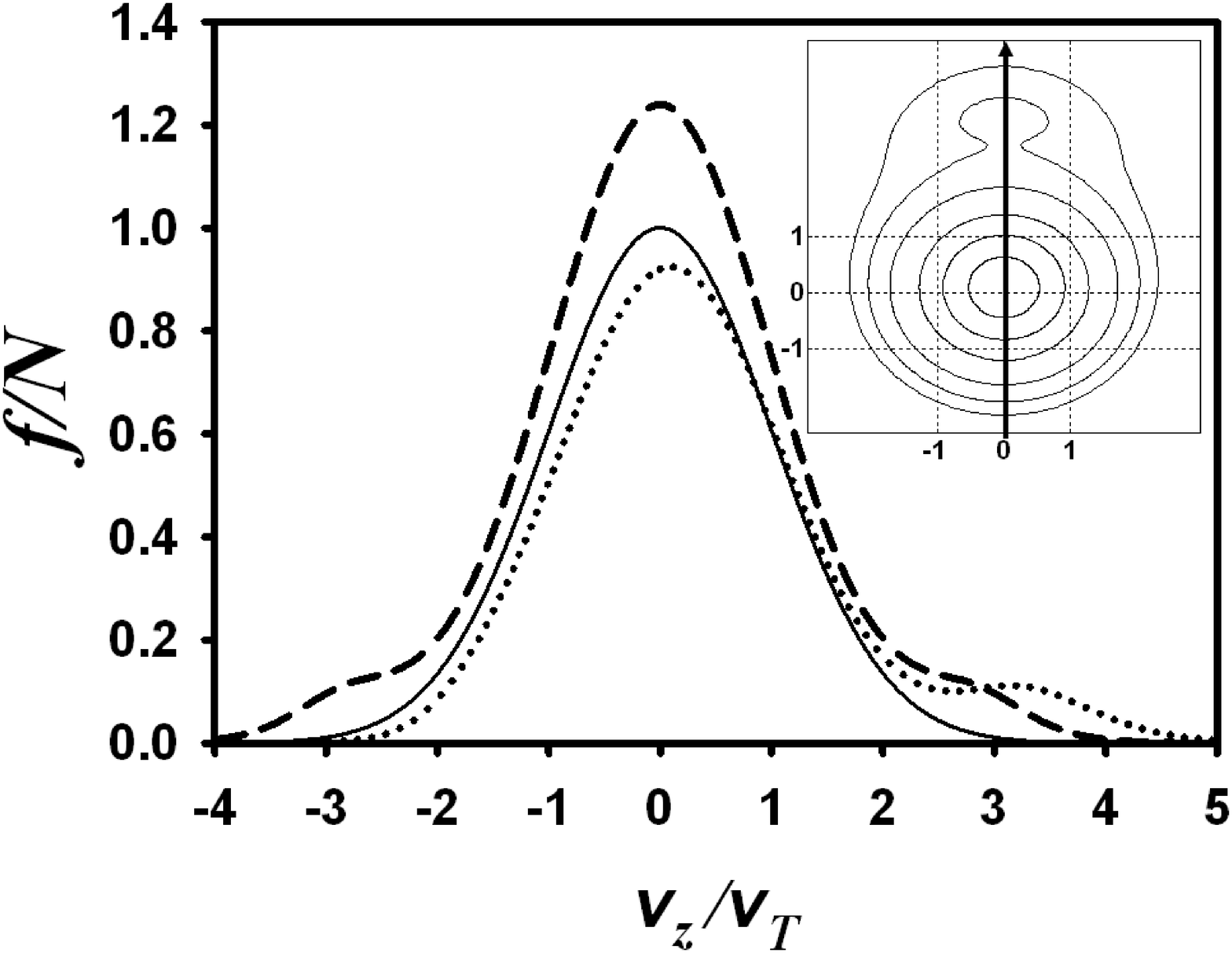}}}
\caption{Distribution as a function of $v$ in current direction (arrow in inset). 
Solid: equilibrium $f=f_{0}$; dashed: $j=0$, $n=2n_{0}$; dotted $n=n_{0}$, $j=n_{0}v_{T}/2$. Inset:
$f$-contour for $n=n_{0}$, $j=n_{0}v_{T}/2$. Tail regions with lower $r(v)$ (weaker equilibration) show stronger
nonequilibrium.}
\label{Fig1}
\end{figure}

{\em Energy Gap, $3d$, Finite Temperature -} Consider now $\mu < 0$, such that $f_{0}\approx N \exp(-mv^2/2kT) \ll 1$,
with $N=\exp (\mu/kT)$, describes a dilute, non-degenerate, equilibrium electron gas in an insulator material.
The terms $-rf$ and $rf_{0}$ describe trapping and emission of thermally equilibrated electrons from electron traps, respectively.
The non-equilibrium distribution is calculated
for a step function-like rate $r(v) = r_{1}$ for $v$ $\ll v_{T}:=\sqrt{kT/m}$ and $r(v)= r_ {2}=0.1\, r_{1}$
for $v$ $\gg v_{T}$, which models a mobility edge (inset in Fig. \ref{Fig2}). The resulting
distribution function $f$ is shown in Fig. \ref{Fig1} for three different cases. Again,
in $v$-regions of larger scattering rates $r(v)$, $f$ is closer to $f_{0}$ because equilibration is stronger.
The effective rate $r^{(j)}(n,j)$, far from equilibrium, as a function of the current density $j$ (with ${\bf j}=j\hat e_{z}$ in $z$-direction)
is shown for different values of $n$
in Fig. \ref{Fig2}. The near equilibrium result, $r^{(j)}(n_{0},j=0) =0.306\, r_{1}$ can be obtained directly from
Eq. (\ref{r}). For large currents,
$j \gg n_{0}v_{T }$, as one expects $r^{(j)}\to r_{2}$. It is clear that an additional consideration of higher order moments will allow
to describe hot electrons.\\ \indent
%
 
\begin{figure}[h]
\centering
    \rotatebox{0}{
        \resizebox{0.4\textwidth}{!}{
            \includegraphics{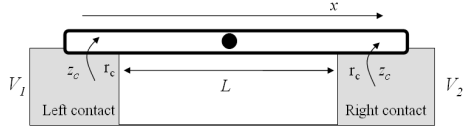}}}
\caption{Quantum wire connected to
two electrodes at temperature $T $ and potentials $V_{1,2}$, with
contact impedance per length, $z_{c}$, and scattering rate $r=r_{c}$.
The ballistic region ($r\to 0$) of length $L$,
contains a localized elastic scatterer with transmission probability ${\cal T}$.}
\label{1DChannel}
\end{figure}

{\em Low frequency admittance of a 1d wire -}
The generalized hydrodynamics framework is very useful to describe low-frequency transport in mesoscopic conductors.
To illustrate this, the admittance of the one-dimensional symmetric wire shown in Fig.\ref{1DChannel} is calculated.
The contacted long end pieces are described by a finite $r\equiv r_{c}$, which models in a natural way
contacts to macroscopic electron reservoirs, where carriers from the wire are absorbed ($-r_{c}f$) and equilibrated carriers
are emitted ($r_{c}f_{0}$) into the wire. For $-L/2<x<L/2$ the electrons are supposed to move ballistic, except
for a localized elastic scatterer at $x=0$ with transmission probability ${\cal T}=1-{\cal R}$. Hence,
$ \mathcal{L}(f_{0}-f) = \nu \delta(x) v_{F}(f(-v)-f(v))$ in $-L/2<x<L/2$ with $\nu={\cal R}/{\cal T}$.\\ \indent
Since the purpose is to illustrate the principle, the discussion is restricted to linear response (cf. \cite{BriggsLeburton1989}).
Coulomb interaction can be included self-consistently \cite{ChristenButtiker1996}. We assume
negligible capacitive coupling between the contacts, small contact impedance
between the substrate and the contacted wire pieces, and obviously disregard electron-electron interaction effects
like Luttinger liquid formation \cite{Yacoby1996} or thermal conductance decrease \cite{Rechetal2009}.\\ \indent
The low-frequency admittance, $G=G_{0}-i\omega E$, for an applied voltage $\Delta V \exp (-i\omega t)$, with $\Delta V = V_{1}-V_{2}$,
includes the DC conductance, $G_{0}$, and the emittance, $E$ \cite{ChristenButtiker1996,Gabelli2007}. 
The linearized hydrodynamic equations (\ref{nh})-(\ref{jh}) can be written as
\begin{eqnarray}
v_{F}\kappa \Delta n + \partial _{x} j & = & 0  \label{n1dl} \\ 
\kappa j + v_{F}\partial _{x}  \Delta n  & = & \frac{2e}{h} \partial_{x} U - 2\nu\delta (x)j \;\; ,\label{j1dl} 
\end{eqnarray}
where $\partial _{x}$ denotes the spatial derivative,
$\Delta n = n-n_{0}$, and $\kappa = (r-i\omega )/v_{F}$ with $r=r(v_{F})$. By virtue of $v_{F} = hn_{0}/2m$, $n_{0}$ was eliminated.  
The applied voltage is equal to the total difference in the electro-chemical potential, 
$V= U-hv_{F} n /2 e$.
The electrical behavior of the wire-support contact is modeled by a phenomenological contact impedance $z_{c}$
with $-e z_{c} \partial_{x}j = V_{1,2}-U(x)$ on the two sides.\\ \indent
In order to derive $G$, we integrate Eq. (\ref{j1dl}) from $x=-\infty$ to $x=\infty$ and obtain with
$\Delta V = V_{1}-V_{2}$, $I(x)=-ej$, and $\kappa _{c}= (r_{c}-i\omega )/v_{F}$:
\begin{equation}
\frac{e^2}{h} \Delta V = \int _{-\infty}^{-L/2} \kappa _{c}I(x) \, dx +(\nu-\frac{i\omega L}{2})I(0),
\label{DeltaVIntegral}
\end{equation}
where the symmetry of the wire was used. The current in the wire in the left contact region $x<-L/2$
must decay, hence $j\propto \exp (\beta x)$ there, with $\beta = (\kappa _{c}^{-2}+
2e^2z_{c}/h\kappa_{c})^{-1/2}$ that follows from Eqs. (\ref{n1dl}) and (\ref{j1dl}).
After evaluation of the integral in Eq. (\ref{DeltaVIntegral}) one obtains 
$e^2 \Delta V /hI(0)= 1+\nu + e^2z_{c}\kappa _{c}/h -i\omega L/2v_{F}$.
With $1+\nu=1/{\cal T}$ and $z_{c}\to 0$, the low frequency admittance $G=I(0)/\Delta V$ becomes
$G= \frac{2e^2}{h}{\cal T}+i\omega \frac{D}{4}{\cal T}^2$
with $D=4e^2L/hv_{F}$, where factors of $2$ for two spin states are included. 
This result is in accordance with Ref. \cite{ChristenButtiker1996} in the considered limit
case of vanishing capacitance between the contacts.
$V(x)$ in the ballistic wire regions has the meaning of a quasi Fermi-level, and the local entropy production rate
$ \dot s(x) = mr(I^{2}+ v_{F} e\Delta n ^2) /T n_{0}e^2$ is localized
in the contact regions.
A generalization to ballistic electrons in arbitrary geometries and arbitrarily far from equilibrium
is straight-forward; it requires higher order
(multipole) moments in $d>1$, similar to a $P-N$ approximation in (non-diffusive) radiation \cite{SiegelHowell1992}. The
generalized hydrodynamic equations with the discussed closure may serve as a general footing for
simulations of nano-electronics devices in the full range between diffusive and ballistic
transport.\\ \indent
%

{\em Conclusion -} It is also straight-forward to extend the method to other than parabolic
energy-momentum relations and to generalized moments \cite{Struchtrup1998}. For instance, it should be
possible to treat in a similar way massless Fermions like neutrinos in stars or electric conduction in graphene,
if these particles are independent and the particle-medium interaction can be modeled by $\mathcal{L}(f_{0}-f)$.\\ \indent
{\em Acknowledgement -} The author thanks Frank Kassubek for his valuable contributions.


\end{document}